\documentclass[superscriptaddress,twocolumn,aps,prb]{revtex4-1}
\usepackage{graphicx}
\usepackage{amssymb,amsmath}
\usepackage{array}
\usepackage{dcolumn}
\usepackage{epsfig}

\usepackage[french,english]{babel}
\usepackage[utf8]{inputenc}
\usepackage{color}
\usepackage[usenames,dvipsnames]{xcolor}

\begin{document}

\title{Electronic structure of the $\rm Ca_3Co_4O_9$ compound from 
ab initio local interactions}
 
\author{Julien Soret}
\author{Marie-Bernadette Lepetit}

\affiliation{CRISMAT, ENSICAEN-CNRS UMR6508, 6~bd. Mar\'echal Juin, 14050 Caen, 
FRANCE}

\date{\today}

\begin{abstract}
  We used fully correlated ab initio calculations to determine the effective
  parameters of Hubbard and t – J models for the thermoelectric misfit
  compound $\rm Ca_3Co_4O_9$. As for the $\rm Na_xCoO_2$ family the Fermi
  level orbitals are the $a_{1g}$ orbitals of the cobalt atoms~; the $e'_g$
  being always lower in energy by more than 240\,meV. The electron correlation
  is found very large $U/t\sim 26$ as well as the parameters fluctuations as a
  function of the structural modulation. The main consequences are a partial
  $a_{1g}$ electrons localization and a fluctuation of the in-plane magnetic
  exchange from AFM to FM. The behavior of the Seebeck coefficient as a
  function of temperature is discussed in view of the ab initio results, as
  well as the 496\,K phase transition.
\end{abstract}


\maketitle

\section{Introduction} \label{intro}

First studied for their ionic conduction properties~\cite{Na-batteries}, the
layered cobalt oxides have been, in the last years, the object of a regained
attention, due to the discovery of peculiar transport and magnetic properties.
One can cite, for instance, the superconducting state of the $\rm Na_{0.35}
CoO_2 - 1.3H_2O$,~\cite{Na-supra} the large thermoelectric power (TEP) in the
$\rm Na_{0.7} CoO_2$,~\cite{Na-TEP} or the rich phases diagram of the $\rm
Na_{x} CoO_2$ family.~\cite{Na-DiagrPh}  In terms of applications the large
thermoelectric power (TEP) found in some materials of this family is certainly
the most interesting property. Indeed, in addition to Seebeck coefficients
larger than $100 \rm \mu V/K$ at room temperature, these systems present the
atypical association of a low resistivity and a low thermal conductivity.  It
results a large figure of merit $ZT=S^2T/ \rho \kappa$ at room
temperature~\cite{Na-ZT,Mf-TEP2} (where $Z$ is the factor of merit,
$S$ the Seebeck coefficient, $T$ the temperature, $\rho$ the electrical
resistivity, $\kappa$ the thermal conductivity).

The motivation was at the origin  the search for new compounds, based on
the same $\rm CoO_2$ layers (known to be responsible for the electronic and
magnetic properties), but with better chemical stability than the $\rm Na_{x}
CoO_2$ ones. Indeed, in air, these systems can both easily accept water or
other small molecules in between the $\rm CoO_2$ layers and loose part of
their sodium content, both processes affecting the desired properties.  The
main idea was to replace the alkaline layer by a more complex oxide one. Many
systems were thus synthesized such as $\rm
Tl_\alpha[(Sr,Ca)_{1-\beta}]_{1+x}CoO_2$,~\cite{Mf1} $\rm
Bi_\alpha[A_{0.75}Bi_{0.25}O]_{3+3x/2}CoO_2$ ($\rm A=Ca, Sr$)~\cite{Mf2},
$\rm Sr_2O_2CoO_2$,~\cite{Mf3} etc.

Among these layered cobalt oxides, the so-called $\rm
Ca_3Co_4O_9$~\cite{Mf-TEP1} ($\rm [Ca_2CoO_3]_{0.62}[CoO_2]$) occupies a
special place. Indeed, this chemically ``simple'' system has been used as the
reference system for the study of the thermoelectric properties in $\rm
CoO_2$-based compounds.  In addition to its interesting thermoelectric
properties the $\rm Ca_3Co_4O_9$ present puzzling magnetic
transitions~\cite{Mf-SDW} with the inset of an incommensurate spin density
wave (IC-SDW) at low temperature, an often called ``ferrimagnetic'' transition
around 19~K and a transition interpreted as a cobalt spin state transition
around 380~K. The long-range IC-SDW sets in place at 27~K however from 100~K a
short range IC-SDW order is observed by muon relaxation~\cite{Mf-SDW}.

\begin{figure}[h]
\centerline{\resizebox{7cm}{!}{\includegraphics{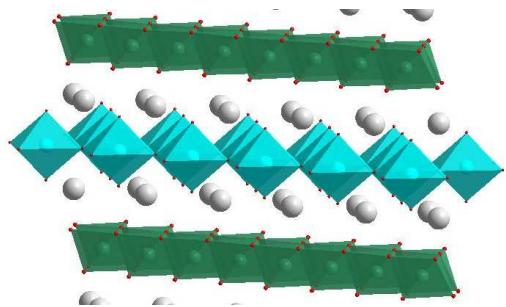}}}
  \caption{(Color online) Structural representation of the $\rm Ca_3 Co_4 O_9$
    compound. Green~: $\rm CoO_2$ layers of edge-sharing octahedra, blue~:
    cobalt octahedra in the rock-salt layers and light gray~: calcium ions.  }
\label{f:struc}
\end{figure}
The $\rm Ca_3Co_4O_9$ system is built from the alternation of $\rm CdI_2$-type
$\rm CoO_2$ layers and $\rm CoCa_2O_3$ rock-salt-type layers~\cite{Mf-Struct}
(see fig.~\ref{f:struc}). These layers are stacked along the $\vec c$-axis and
present incommensurate lattice parameters along the $\vec b$ direction.  The
$\rm CoO_2$ layers are built from edge-sharing cobalt octahedra presenting a
compression along the $\vec c$ crystallographic direction (the octahedra (111)
direction). The rock-salt layers are three-fold layers built from two Ca-O
planes sandwiching a Co-O plane, or equivalently built from a layer of cobalt
octahedra, as in a rock-salt structure. The
incommensurability between the $\vec b$ lattice parameters of the two
subsystems induces, in each layer, a structural modulation presenting the
periodicity of the other subsystem. One of the important consequences is a
distortion of the coordination sphere of each cobalt atom in the $\rm CoO_2$
layers (different Co-Co distances, different Co-O distances and Co-O-Co
angles).  Indeed, the $\rm CoO_2$ subsystem has been shown to be responsible
for the low energy properties of the compound (transport, TEP, magnetism,
etc.), since it supports the Fermi level electrons~\cite{Mf-TEP1,Mf-TEP2}
while the $\rm CoCa_2O_3$ subsystem was proved to be gapped. In addition, in
the related $\rm Na_xCoO_2$ family the Fermi level properties were proved to
be strongly dependant both on the doping~\cite{Na-DiagrPh} and on the local
structural parameters~\cite{CoO2_3}.

The purpose of the present work will thus be to study the effect of the
incommensurate modulations on the low energy degrees of freedom of the $\rm
Ca_3Co_4O_9$ system.  At this point one should notice that, unfortunately,
density functional (DFT) electronic structure calculations fail to properly
describe the electronic structure of the $\rm CoO_2$-based systems.  Indeed,
LDA as well as LDA+U calculations~\cite{Mf-dft} exhibit a conducting $\rm
CoCa_2O_3$ subsystem dominating the density of states at the Fermi level. DFT
is known to illy treat strong correlation effects, however, in many strongly
correlated systems, it is nevertheless able to yield reasonable Fermi surfaces
and the correct magnetic ground state. In the $\rm CoO_2$-based systems, this
is not the case. In the $\rm Na_xCoO_2$ compounds DFT calculations yield a
ferromagnetic ground state~\cite{Na-dft1}, while experimentally
antiferromagnetic~\cite{Na-AFM}, and incorrect Fermi
surfaces~\cite{Na-dft2}. We will thus need to rely on alternating approaches,
such as embedded cluster quantum chemical methods, explicitely treating the
strong correlation at the Fermi level. Such methods also present the
advantages to be able to treat the misfit character of the system.
For this purpose we will use the same ab initio
method that gave accurate and reliable results on the evaluation of such local
effective parameters in other strongly correlated systems, that is the
embedded fragment CAS+DDCI method~\cite{method}. This method yielded very good
results in systems such as the copper or vanadium oxides~\cite{DDCI_ex}, the
$\rm Na_x CoO_2$~\cite{CoO2_1,CoO2_3}, etc., in periodic systems as well as
misfit ones~\cite{numtel}.

The modulated structures will be taken from
reference~\onlinecite{Mf-Struct}. We will compute the magnetic effective
exchange between two nearest neighbors (NN) cobalt ions, the effective
transfer, the correlation strength and the on-site parameters variations that
could be responsible for electron localization.

\section{Method and computational details} \label{methode}

The embedded fragment CAS+DDCI method is a spectroscopy method specifically
designed to accurately treat the strongly correlated character of a set of
orbitals (designed as active or magnetic orbitals), as well as the screening
effects on all processes on this set of magnetic orbitals. Such a calculation
implies large diagonalizations and can only be done on a formally finite
system. 
To reach the desired goals, the diagonalized
  configurations space includes (1) all configurations that can be built
  within the magnetic orbitals (Complete Active Space or CAS), (2) all single
  and double excitations on each of the previous configurations, except for
  the double excitations not contributing to the excitations energies in a
  perturbative analysis. 
Such a choice insures a proper treatment of the
correlation effect on the cobalt $3d$ shell as well as the screening effects
on the cobalt $3d$ electrons~\cite{method}.

The embedded fragments are thus composed of a quantum part, treated within the
CAS+DDCI quantum chemical method, and an embedding that reproduces the main
effects of the rest of the crystal on the quantum part.  The quantum part
should include the cobalt ions involved in the sought interaction and their
first coordination shell. 

The main effects of the rest of the crystal on the quantum part are
double. First, there is the short range exclusion of the quantum electrons
from the space occupied by the electrons of the rest of the crystal. This
effect is modeled using total ions pseudo-potentials~\cite{TIPs} (TIPs) that
represent the electronic structure of the first shells of atoms surrounding
the quantum part. The second important effect is the Madelung potential. For
this purpose, we used a set of renormalized charges. As far as
the two periodic directions are concerned, we used the real space method
developed by A. Gell\'e {\it et al.}~\cite{env2} that reproduces the
Madelung potential with an exponential
convergence. For the incommensurate direction, we used the same method as in
reference~\onlinecite{numtel}, imposing the nullity of both charge and dipole
moment. 

The question is now {\em ''what formal charges should we use for the different
  atoms 
  in order to build the embedding
  ~?''}. The calcium and oxygen atoms will be taken as $\rm Ca^{2+}$ and $\rm
O^{2-}$ ions. The question of the cobalt valence is a bit more
difficult. Indeed, there is no reason for the cobalts to present the same
valency in the two subsystems. Two arguments can help us in finding a
reasonable evaluation of the charge transfer between the two layers. The first
argument is a simple comparison between the average first neighbor
cobalt--oxygen distances in the $\rm CoO_2$ layers for $\rm Ca_3Co_4O_9$ and
for the $\rm Na_xCoO_2$ family. With an average Co--O distance of $1.904\rm
\AA$ the cobalt valence could be estimated from figure~5 of
reference~\onlinecite{Na-dist} to be similar to the $\rm Na_{2/3}CoO_2$
system, that is about $\rm Co^{3.33+}$ in the $\rm CoO_2$ layer, and thus $\rm
Co^{3.05+}$ in the rock-salt layer. The second argument comes from a recent
evaluation of the cobalt valency using atomic-column resolved energy-loss
spectroscopy (EELS)~\cite{Mf-EELS}. Indeed Klie {\it et al} found a single
valency of $\rm Co^{3+}$ in the rock-salt layers and a mixed valency in the
$\rm CoO_2$ layers. We thus used a $\rm Co^{3+}$ formal charge for the
rock-salt subsystem and, insuring electro-neutrality, a $\rm Co^{3.38}$ formal
charge for the $\rm CoO_2$ subsystem. 
Such evaluations of the cobalts' valences correspond to average values within
each layer and, at this point, one could wonder about the effects of the
fluctuations around these values, on the Madelung potential acting on the
fragments. As a matter of example let us evaluate the order of magnitude of
the potential modification due to an oxygen vacancy in the rock-salt layer,
located in the close vicinity ($\sim$10\AA{}) of the fragment (located in the
$\rm CoO_2$ layer).  The relative modification of the Madelung potential, at
an atomic position within the quantum fragment, scales as $\delta V/V \simeq
-2/R_V/V \sim 10^{-5}$. The relative fluctuation of the potential gradient
(the important degree of freedom for our calculations) is even weaker. In
addition, the later should be partly compensated by a modification of the
valency of the cobalt atoms, located in the vicinity of the vacancy within the
rock-salt layer.  One can thus expect that such fluctuations, provided their
randomness and scarcity will little affect the reliability of our results.

In the quantum calculations, the core electrons where treated using effective
core potentials~\cite{bases}, while the valence and semi-valence electrons
where treated using the associated Gaussian atomic basis set~\cite{bases}.

\section{Results~: the cobalt $3d$  orbitals}   \label{atome}

In the $\rm CoO_2$ subsystem of the $\rm Ca_3Co_4O_9$ compound, the cobalt
ions are in a mixed valence state $\rm Co^{3+}$ ($3d^6$) and $\rm Co^{4+}$
($3d^5$). It is sometimes written in the literature~\cite{Mf-TEP1,Mf-Struct2}
that the $\rm Co^{3+}$ ions are in a low spin state while the $\rm Co^{4+}$
ions are in a high spin state.  This assumption is in contradiction with what
was found for the $\rm Na_xCoO_2$ family where both cobalt ions are in a low
spin state ---~$\rm Co^{3+}~:~ t_{2g}^6e_g^0$ and $\rm Co^{4+}~:~
t_{2g}^5e_g^0$.  Since these $3d$ orbitals are responsible for the physical
properties, it is of prime importance to first understand the cobalt $3d$
orbital splitting and spin states.

We thus computed the different spin states of a $\rm CoO_6$ embedded fragment
($\rm Co^{3+}$ as well as $\rm Co^{4+}$ ionic states) for all
crystallographically independent cobalt sites in the periodic directions, and
a representative set of them in the misfit direction. We found that in all
cases both the $\rm Co^{3+}$ and $\rm Co^{4+}$ ions are in a low spin
state. Indeed, even for the $\rm Co^{4+}$ valence state the higher spin states
are at least at 700~meV ($\sim 8000\,$K) above the low spin state.

As in the $\rm Na_xCoO_2$ family, the $\rm CoO_6$ octahedra are compressed
along the $\vec c$ axis. One thus expects a splitting of the regular
octahedron $t_{2g}$ orbitals into a doublet $e_{g}^\prime$ and a singlet
$a_{1g}$. The energetic order between these orbitals is crucial since it
determines the nature of the magnetic orbitals of the $\rm Co^{4+}$ ions.  A
simple crystal field evaluation yields the $e_{g}^\prime$ orbitals at the
Fermi level. However, it was shown that due to the $O_h$ to $S_6$ symmetry
reduction, the $e_{g}^\prime$ orbitals can and do hybridize with the high
energy $e_g$ ones. It results an energetic stabilization of the $e_g^\prime$
orbitals compared to the $a_{1g}$ one and thus a non degenerated atomic ground
state for the $\rm Co^{4+}$ ions~\cite{CoO2_2}~: $(e_g^\prime)^4a_{1g}$. In
the present system the misfit character of the two subsystems tells us that
further symmetry breaking occurs and that all cobalt $3d$ orbitals are non
degenerated.

\begin{figure}[h]
\centerline{\resizebox{7cm}{!}{\includegraphics{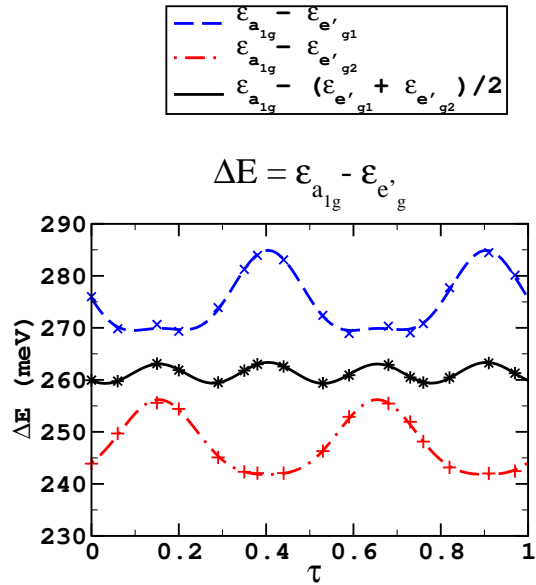}}}
\caption{(Color online) $3d$ cobalt orbital energy splittings~:
  $\varepsilon_{a_{1g}}-\varepsilon_{e^\prime_g}$, as a function of the
  incommensurate modulation parameter $\tau$.  Blue curve with crosses~:
  $\varepsilon_{a_{1g}}-\varepsilon_{e^\prime_{g1}}$ splitting, red curve with
  plus signs~: $\varepsilon_{a_{1g}}-\varepsilon_{e^\prime_{g2}}$ splitting,
  black curve with stars~: average  $\varepsilon_{a_{1g}}-\varepsilon_{e^\prime_g}$ splitting.}
\label{fig:split}
\end{figure}

Figure~\ref{fig:split} displays the
$\varepsilon_{a_{1g}}-\varepsilon_{e_{g}^\prime}$ orbital energy splitting as
a function of the fourth crystallographic dimension parameter ($\tau=mod(n
\,b_{\rm CoO_2}/b_{RS},1), \, n\in {\mathbb N}$ refers to the $\rm CoO_2$
subsystem cells in the $\vec b$ direction ---~see figure~\ref{fig:dimer}~---
$b_{\rm CoO_2}$ and $b_{RS}$ are the lattice parameters of the two
subsystems in the incommensurate direction) associated with the
incommensurate modulations of the $\rm CoO_2$ layers. Let us recall that this
structural modulation takes place along the $\vec b$ direction.  These
effective orbital splittings where obtained as the energy difference of fully
correlated ${e_g^\prime}^4 {a_{1g}}^1$ and ${e_g^\prime}^3 {a_{1g}}^2$ states
of the $\rm CoO_6$ embedded fragments. Indeed, this excitation energy can
be associated with the effective orbital splitting of a 3-bands Hubbard model.
One sees immediately that whatever the Co site, the $a_{1g}$ orbital is always
of higher energy than the $e^\prime_{g}$ ones~; that is the hole is always
located in the $a_{1g}$ orbital. The excitation energy for having a hole
located in a $e^\prime_{g}$ orbital is always larger than $240\,{\rm
  meV}\simeq 2800\,{\rm K}$. The average
$\varepsilon_{a_{1g}}-\varepsilon_{e^\prime_g}$ splitting is 261\,meV, while
the difference between the two $e^\prime_{g}$ orbitals ranges between 13\,meV
to 42\,meV. Finally one should notice that the variations of the
$\varepsilon_{a_{1g}}-\varepsilon_{e^\prime_{g}}$ splitting as a function of
the different crystallographic sites remains small ($<\pm 10\,$meV).

As mentioned above the relative position of the $a_{1g}$ and $e^\prime_{g}$
orbitals are related to the $e^\prime_{g}$--$e_g$ hybridization. As expected
from the weak modulation of the orbital energy splitting, this hybridization
remains stable, close for all sites to its average value of $10$ degrees. This
value is very comparable to the values found on different $\rm Na_x CoO_2$
systems~\cite{CoO2_3}.

\section{Results~: $a_{1g}$-based one band models}   
 \label{modele}

The previous section clearly told us that the $a_{1g}$--$e^\prime_{g}$ orbital
splitting is large enough ($2800\,{\rm K}$) to justify a on-band model for our
system. In this section, we will thus extract from our calculations, both the
effective exchange integrals, and the parameters of an extended Hubbard model.

The NN effective exchange can be obtained as the energy difference between the
singlet and triplet states of the $\rm Co_2O_{10}$ embedded fragments. The
extended Hubbard model parameters necessitate in addition the states wave
functions information. Indeed, the parameters are determined so that the model
reproduce both the singlet--triplet excitation energy, and the relative weight
of the dominant configurations, in the states wave functions. One should
however notice that, not all the parameters of the extended Hubbard model, can
be independently determined from the data. Indeed, only the NN hopping
integral~: $t$, the difference between the average on-site repulsion and the
NN repulsion~: $\bar U-V=\left(U_1+U_2\right)/2-V_{12}$ and the difference
between the orbital energies and the on-site repulsions~: $\delta
\varepsilon_{a{1g}} - \delta U$ can be independently determined.

In this non-periodic system, the models  parameters were determined for a
representative set (sixteen dimers were computed) of incommensurate
distortions values, then fitted as a function of the  incommensurate dimension
parameter $\tau$. 
\begin{figure}[h]
\resizebox{6cm}{!}{\includegraphics{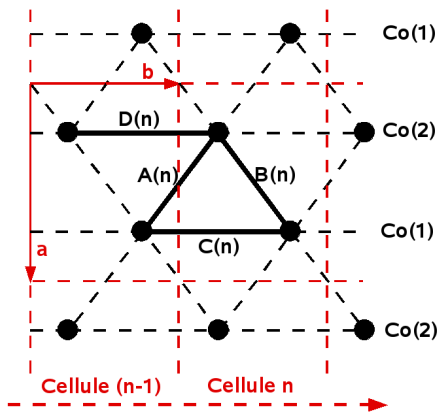}} \vspace*{0.5eM}
\caption{(Color online) Schematic representation of the different dimers.}
\label{fig:dimer}
\end{figure}

One can define two types of non equivalent nearest neighbor dimers
that propagate along the incommensurate $\vec b$ direction (see
figure~\ref{fig:dimer}); the A dimers ($\vec{a}-\vec{b}$ direction) and
the C dimers ($\vec{b}$ direction). Indeed, the B and D dimers can be
deduced from the preceeding ones, through the expressions:
\begin{align}
  \tau_A & \rightarrow \tau_B=0.5\frac{b_{\rm CoO_2}}{b_{RS}}-\tau_A \\ 
  \tau_C & \rightarrow \tau_D=-\tau_C
\end{align}
In the following discussions, we will refer to ``$\vec{b}$ direction dimers''
for the C and D dimers and to ``$\vec{a}\pm\vec{b}$ directions dimers'' for
the A and B ones.

\subsection{Effective exchange and hopping}

\begin{figure}[h]
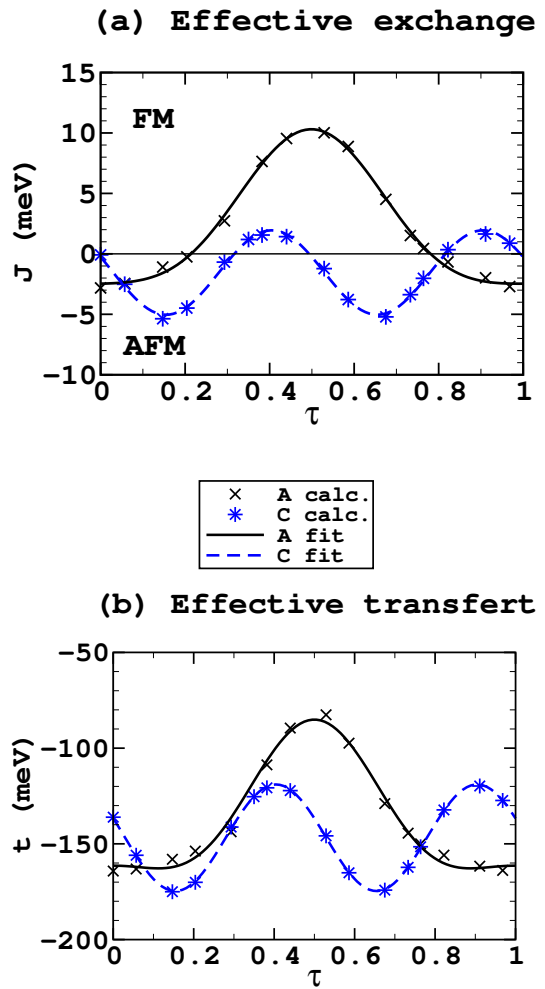

\resizebox{7cm}{!}{\includegraphics{Ca3Co4O9.J.tau.eps}} \\[2eM]
\resizebox{7cm}{!}{\includegraphics{Ca3Co4O9.t.tau.eps}}
\caption{(Color online) (a) Effective exchange, $J$, and (b) effective
  transfer, $t$, as a function of the fourth crystallographic dimension
  parameter $\tau$. Symbols represent the computed values, lines the fits.}
\label{fig:J_tau}
\end{figure}

Figure~\ref{fig:J_tau}(a) reports the variations of the effective exchange
integral $J$ and figure~\ref{fig:J_tau}(b) of the effective hopping integral
$t$, as a function of the fourth crystallographic dimension parameter
$\tau$. Table~\ref{table:param} reports the parameters for the fit of the
computed values according to the formula
\begin{eqnarray} \label{eq:fit} 
J(\tau)&=&J_0 \,+\, J_1\cos{\left(2\pi[\tau+\varphi]\right)}
              \,+\, J_2\cos{\left(4\pi[\tau+\varphi]\right)}
\nonumber \\
t(\tau)&=&t_0 \,+\, t_1\cos{\left(2\pi[\tau+\xi]\right)} 
              \,+\, t_2\cos{\left(4\pi[\tau+\xi]\right)}   
\end{eqnarray}

\begin{table}[h]
\begin{center}
  \begin{tabular}{l|cccc|cccc} \hline
    Dimer & $J_{0}$ & $J_1$ & $J_2$& $\varphi$ & $t_0$ &  $t_1$ & $t_2$ & $\xi$ \\
    \hline
    A  &  2.56 & -6.38 & 1.36 & 0.00 & -135.81 & -38.15 & 12.65 & 0.00 \\  
    C  & -1.54 & -0.02 & 3.49 & 0.60 & -146.82 &   0.11 & 27.85 & 0.10 \\ 
    \hline
  \end{tabular}
\end{center}
\caption{Parameters for effective exchange $J$ and effective
  transfer $t$ fits as a function of the incommensurate dimension parameter
  $\tau$. The fitted values are associated with equations~\ref{eq:fit}.}
\label{table:param}
\end{table}
Let us notice  the following points.
\begin{itemize}
\item 
  The NN exchange $J$ may be either antiferromagnetic (AFM) or ferromagnetic
  (FM) (resp. negative and positive) according to the specific dimer.
\item 
  The NN exchange and hopping are strongly modulated for all dimer
  types. These modulations are essentially related to the Co-O distance
  modulations and not to the Co-Co ones. Indeed, despite the fact the Co-Co
  distances are nearly constant for the C and D dimers ($\vec{b}$ direction
  dimers), and strongly fluctuating for the A and B ones ($\vec{a}\pm\vec{b}$
  directions dimers)~\cite{Mf-Struct}, the variation range of both exchanges and
  hoppings present similar amplitudes ($\sim$7\,meV and $\sim$13\,meV
  respectively). In fact, one should notice that the amplitude of both the
  super-exchange AFM contribution to $J$, and the hopping integral, decrease
  when the Co-O distance increases.
\item 
  $J$ absolute values are very small in amplitude, with average values
  around 2.56\,meV for $\vec{b}$ direction dimers, and -1.54\,meV for
  $\vec{a}\pm\vec{b}$ direction dimers (see
  Table~\ref{table:param}). The NN exchange is thus mostly
  ferromagnetic along the $\vec{b}$ direction and antiferromagnetic
  along the $\vec{a}\pm\vec{b}$ directions.
\end{itemize}

\begin{table}[h]   
  \begin{center}
    \begin{tabular}{c|ccc|c} \hline
      System  & \multicolumn{3}{|c|}{$\rm Na_x CoO_2$} & $\rm Ca_3Co_4O_9$
      \\ \hline
      $x$     & ~0~ &   ~0.35~ & 0.5 & $\sim 0.62$ \\
      $J$ (meV) &  -52~\cite{CoO2_3} &  -36~\cite{CoO2_1} & -11/-19/-19/-27~\cite{CoO2_3} & [-2.5,10.3]\,/\,[-5.0,2.0] \\  \hline
    \end{tabular}
  \end{center}
  \caption{Effective exchange values (in meV) as a function of the $\rm CoO_2$
    filing parameter $x$, present system compared to different compounds of 
    the  $\rm Na_x CoO_2$ family.}
  \label{table:J}
\end{table}
Comparing these exchange values with the ones obtained in the $\rm Na_x CoO_2$
family, one sees that while for the superconducting system~\cite{CoO2_1}
($x=0.35$), the $x=0.5$ and the $x=0$ systems~\cite{CoO2_3} the exchange are
always AFM, in the present system both AFM and FM interactions were found
according to the Co--Co bond. In fact the antiferromagnetic character of $J$
decreases with increasing $x$ (see table~\ref{table:J}) for the $\rm Na_x
CoO_2$ compounds. These theoretical findings are in agreement with
experimental data. Indeed, for low $x$ values, the $\rm CoO_2$ layers of the
$\rm Na_x CoO_2$ systems were found to be anti-ferromagnetically
coupled~\cite{Na-AFM}, while for large $x$ values, neutrons scattering exhibit
A-type antiferromagnetism~\cite{Na-FM}, that is ferromagnetic
correlations within the $\rm CoO_2$ layers which are anti-ferromagnetically
coupled. In fact, Lang {\it et al} NMR studies~\cite{Na-DiagrPh} allowed to
propose a magnetic phases diagram with a temperature dependant boundary
between the AFM and FM correlations within the $\rm CoO_2$ layers. At low
temperature the critical doping $x*$ is in the $0.6\le x^*\le 0.7$ range and
decreases with increasing temperature ($x^*(300\,\rm{K})\sim 0.5$). Comparing
now our computed $J$ values for the $\rm Ca_3Co_4O_9$ compound with both
theoretical and experimental results for the $\rm Na_x CoO_2$ family, one sees
that the present compound ---~associated with $x\sim 0.62$~--- compares well
with the $\rm Na_x CoO_2$ family. Indeed, it fluctuating FM/AFM effective
exchange agrees well with its assumed doping of the $\rm CoO_2$ layers close
to the phase boundary.

\subsection{Effective on-site repulsion}
\begin{figure}[h]
  \centering
\centerline{
\resizebox{7cm}{!}{\includegraphics{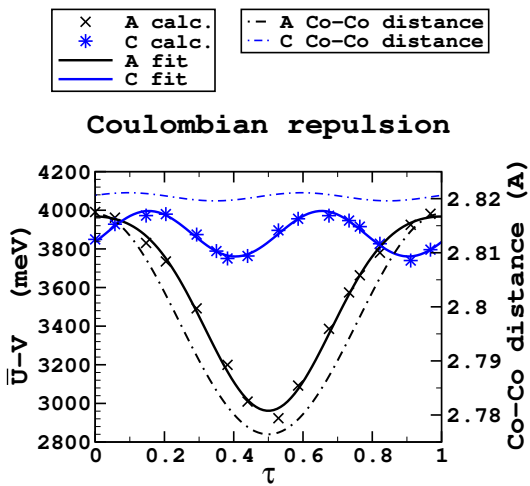}}}
\caption{(Color online) Left scale~: difference (meV) between the average
  on-site repulsion on a dimer $\langle U \rangle$ and the effective NN
  repulsion $V$ as a function of the fourth crystallographic dimension
  parameter $\tau$. Right scale~: Co--Co NN distances (\AA{}) as a function of
  $\tau$.}
  \label{fig:U}
\end{figure}

Figure~\ref{fig:U} reports the difference between the average on-site
repulsion on a dimer, $\bar U $, and the NN repulsion $V$: $\bar
U-V=\left(U_1+U_2\right)/2-V_{12}$.  The structural analysis shows us that the
Co--Co distance variation is very weak for the $\vec{b}$ direction dimers,
while it is large for the $\vec{a}\pm\vec{b}$ dimers (See
figure~\ref{fig:U}). Since the NN repulsion is essentially related to the
Co--Co distances one can assume that $V$ is nearly constant along the
$\vec{b}$ direction, but strongly varies along the $\vec{a}\pm\vec{b}$
one. Let us analyze figure~\ref{fig:U} in this light.  The $\bar U-V$ curve
exhibits only weak modulations for the C dimers.  Since both $V$ and $U$ can
be assumed as nearly constant for these dimers, the observed modulations may be
related to either of them. In any case one can conclude that $U$ remains large
and nearly constant from site to site (fluctuations range smaller than
$0.2\,\rm eV$ to be compared to an average $\bar U -V$ of 3.88\,eV).  The
sites along the A, B and C, D dimers being the same, it results that the
strong variation observed on the $\vec{a}\pm\vec{b}$ direction dimers is
essentially due to NN repulsion variations $V(\tau)$ associated with the
strong fluctuations of the Co--Co distances. The amplitude of the variation is
quite large, with about 1eV for an average value of 3.5eV (see
table~\ref{table:U-V}). It is not possible using our data to quantify further
the NN repulsion, however our results suggest that $V$ is large (a few eV in
amplitude). An accurate simple model should thus
take into account the NN repulsion between the $a_{1g}$ orbitals and its
strong variations~; the on-site repulsion being taken as site-independent.
The average values $\bar U-V$ values found in the present work are consistent
with the values found for the $\rm Na_{x}CoO_2$ family. Indeed the
super-conducting compound exhibits $U-V=3.6\,\rm eV$ while it was found for
the $\rm Na_{0.5}CoO_2$ system $2.6\,\rm eV$ and $2.8\,\rm eV$ according to
the crystallographic site.

\begin{table}[h]
  \begin{center}
    \begin{tabular}{c|cccc}
      \hline
      Dimer & $a_0$ & $a_1$ & $a_2$& $\zeta$ \\
      \hline
      A & 3542 & 503 & -78 & 0 \\
      C & 3878 & 0   & -118 & 0.1 \\
      \hline
    \end{tabular}
  \end{center}
  \caption{Parameters for $\bar U-V$ fits according to the equation 
    $\bar U(\tau)-V(\tau)=a_0 \,+\, a_1\cos{\left(2\pi[\tau+\zeta]\right)}
    \,+\, a_2\cos{\left(4\pi[\tau+\zeta]\right)}$, where $\tau$ is the fourth 
    crystallographic dimension parameter. }
  \label{table:U-V}
\end{table}

\subsection{Effective $a_{1g}$ orbital energy}
\begin{figure}[h]
\centerline{
\resizebox{7cm}{!}{\includegraphics{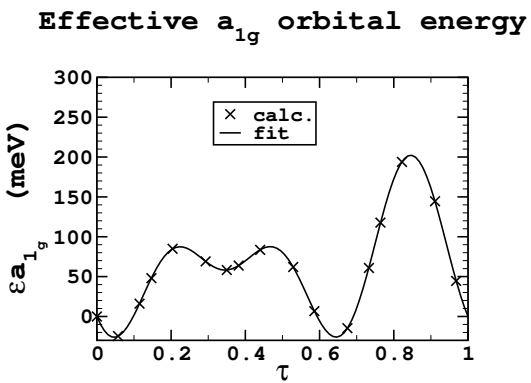}}}
   \caption{$a_{1g}$ orbital energy (meV) as a function of $\tau$. The $\tau=0$
     value is taken as the energy reference.}
  \label{fig:delta}
\end{figure}

The on-site repulsion being considered constant, the energy difference between
NN $a_{1g}$ orbitals can be extracted from our results.
Figure~\ref{fig:delta} reports the effective $a_{1g}$ orbital energy~:
$\varepsilon_{a_{1g}}(\tau)$ ($\varepsilon_{a_{1g}}(0)$ being taken as the energy
reference).
One can see that the amplitude of the $\varepsilon_{a_{1g}}(\tau)$
modulations is quite large ($>200$\,meV). Compared to this modulation,
the $a_{1g}$--$e'_g$ energy splitting is negligible with an amplitude
10 times weaker. It results that the $\varepsilon_{a_{1g}}(\tau)$
modulation can be seen as an effective potential on the whole cobalt
atom.  Associated with the strong NN repulsion V~\cite{V-Hubbard}, one
can thus expect a partial charge localization, with $\rm Co^{3+}$ ions
on the low energy sites.  This result agrees with the observations on
the $\rm Na_xCoO_2$ family for which a partial charge localization is
observed for $x\ge 0.5$~\cite{Na-loc}.

\section{Discussion} \label{discussion}
Let us now take a closer look on the Seebeck coefficient in the $\rm
Ca_3Co_4O_9$ system.  The first point to be stressed is the fact that the TEP
can be quite different according to the intercalation layer ($S^{300K}(\rm
Na_{0.7} CoO_2)\simeq 100 \mu V/K$~\cite{Na-TEP}, $S^{300K}(\rm
[Ca_2CoO_3]_{0.62}[CoO_2])\simeq 120 \mu V/K$~\cite{Mf-TEP1,Mf-TEP2},
$S^{300K}(\rm [CaBiO_2]_2[CoO_2]_{1.69})\simeq 160 \mu V/K$~\cite{Mf-doumerc}).
The second point is the shape of the Seebeck coefficient as a function of
temperature. At low temperature the Seebeck coefficient increases
quasi-linearly up to $\sim$100\,K and then saturates to a constant value up to
about $\sim$400\,K~\cite{Mf-400K}. These two behaviors can be fitted on a
correlated Fermi liquid behavior, for the low temperature part~\cite{Mf-TEP-bt}
($T<100\rm ~K$), and on a Heikes type of configurational entropy for the
saturating part~\cite{Mf-TEP-Hk} (150\,K$\leq T \leq $ 400\,K).

The generalized Heikes formula~\cite{TEPDoumerc} is a high temperature
approximation. It supposes that the system is at very large temperature
compared to the scale of the low energy degrees of freedom. In the present
compound we showed (see section~\ref{atome}) that these degrees of freedom are
associated with the $a_{1g}$ orbital of the $\rm CoO_2$ layers cobalt atoms
$3d$-shell. Using this formula and the non-degenerated character of the
magnetic $a_{1g}$ orbital, one finds (using a cobalt valency of 3.38 as
discussed previously in section~\ref{methode}), a TEP value of $125\,\rm \mu
V/K$. This value is in good agreement with the experimental value of $120\,\rm
\mu V/K$ found for the $\rm Ca3Co_4O_9$ saturation
plateau~\cite{Mf-TEP1,Mf-TEP2}. One may conclude from this agreement
that, at room temperature, the high temperature limit of the triangular one
band model based on the $a_{1g}$ orbitals is already reached.  However, room
temperature only corresponds to 25\,meV, while the band width of a tight
binding model on a triangular lattice is $W=9|t|$~; using our present transfer
estimations ($|t|\ge 100\,\rm meV$) it comes a band width of the order of
10000\,K. How can we account for such a large difference in order of
magnitude? Why does the Heikes model fits so nicely the room temperature
Seebeck coefficient for the $\rm CoO_2$-based compounds?

\begin{figure}[h] \vspace{3ex}
  \centerline{\resizebox{5.5cm}{!}{\includegraphics{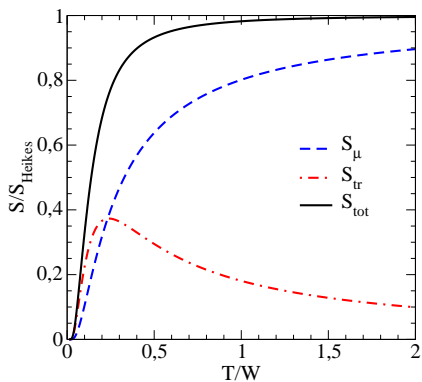}}}
  \caption{(Color online) Seebeck coefficient of a triangular lattice as a
    function of the temperature, normalized to the band width. Blue dashed
    curve~: the entropic term~: $S_{\mu}(T)$, red dot-dashed curve~: the Fermi
    liquid part~: $S_{tr}(T)$.}
  \label{fig:seebeck}
\end{figure}

Different arguments can be advanced in order to explain this high temperature
behavior of the TEP at 300\,K.  \\
Let us first analyze the behavior of the Seebeck coefficient, as a function of
temperature within a one-band tight binding model on a triangular lattice. The
Boltzmann equation yields the following Seebeck expression 
\begin{align}
  S(T) & = -\frac{e}{k_BT}\frac{\int (\varepsilon - \mu) \frac{\partial
      f_0}{\partial \varepsilon} d\varepsilon}{\int \frac{\partial
      f_0}{\partial \varepsilon}
    d\varepsilon} \\
  & = -\frac{e}{k_BT}\frac{\int \varepsilon \frac{\partial f_0}{\partial
      \varepsilon} d\varepsilon}{\int \frac{\partial f_0}{\partial
      \varepsilon}
    d\varepsilon} + \frac{e}{k_BT}\mu\\
  & = S_{tr}(T) + S_{\mu}(T) \\
\end{align}
where $f_0(T,\varepsilon)$ is the Fermi function, $\varepsilon$ the
orbitals energy, $\mu$ the electrochemical potential~; the collision
time as well as the degeneracy is supposed energy
independent. Figure~\ref{fig:seebeck} reports $S(T)$ and its two
components $S_{tr}(T)$ and $S_{\mu}(T)$. At low temperature
$S_{tr}(T)$ dominates, reaches a maximum for $T/W\sim 1/4$, then
slowly decreases toward zero in the high temperature limit. On the
contrary the entropic part $S_{\mu}(T)$ increases for all temperatures
and reaches its high temperature limit for temperatures of the order
of a few band widths. The sum of the two terms results in a Seebeck
coefficient reaching its saturation value at much lower temperature
than its entropic part. Indeed for $T/W=0.5$, $S$ is already larger
than 90\% of its high temperature value. This simple analysis already
allows us to reduce, from a few band widths to half the band width, the
estimation of the temperature at which the TEP saturation should be
observed.  This is however not enough to account for the observed
saturation temperature. Indeed $W/2 = 9|t|/2 $ still accounts for
about 5000\,K. \\
 A second argument is the extremely large value of the correlation
strength in these compounds. Indeed, our calculations yield for the
average $U/t$ a value of 26. Despite this very large $U/t$, transport
measurements~\cite{Mf-TEP1,Mf-TEP2} find the system metallic. One thus
expect its Fermi level quasi-particle pic to be strongly renormalized
due to the strong correlation effects~\cite{Mott}. \\ Finally let us
remember that our results suggest a partial electron localization due
the large site energy modulation ($>200\,\rm meV$) and a large value
of NN repulsion. It results that the lattice pattern seen by the Fermi
level electrons can be expected to be quite different from the
homogeneous triangular lattice assumed in these systems. This effect
is also expected to strongly reduce the quasi-particle bandwidth at
the Fermi level.\\ All together, it seems quite possible that the
bandwidth reduction is large enough (20 to 30 times) so that
$0.5W_\text{eff} \sim 200\,\rm K$, and thus that the high temperature
Heikes model appears valid for the present (and related) systems.

Let us now take a look on the TEP behavior at temperatures larger than
400\,K.  Indeed, at $T_c=396\,\rm K$ the $\rm Ca_3Co_4O_9$ compound
experiences a first-order metal-semiconductor phase transition
associated with a linear increase of the Seebeck coefficient for
$T>T_c$~\cite{Mf-400K}. Different hypotheses have been proposed in
order to account for this phase transition and the high temperature
phase properties. The first one associates the transition with a
cobalt spin-state transition~\cite{Mf-TEP1}. However our calculations
show that the first atomic excited spin state of the cobalt atom is at
least at 700\,meV, i.e. 8100\,K, above the ground state. Such an
energy difference is clearly incompatible with a phase transition
energy scale of 400\,K.  The second hypothesis associates the
transition with a change in the cobalt $t_{2g}$ orbitals
occupancies. However this hypothesis does not resist to the energy
scale of the $a_{1g}$--$e'_g$ energy splitting found in
section~\ref{atome}. Indeed, the 240\,meV$\simeq$2800\,K is still far
too large for the 400\,K energy scale of the transition. The last
hypothesis is an electron localization of the $e'_g$ orbitals,
associated with a structural phase transition~\cite{Mf-400K}. Even if
the assumption that the $e'_g$ orbitals are the Fermi level ones has
been proven incorrect (see reference~\onlinecite{CoO2_3} or for the
present system section~\ref{atome}), the idea of an electron
localization induced by an increase of the in-plane distances in the
$\rm CoO_2$ layers seem quite possible and would explain the
metal-semi-conductor phase transition. Muguera and
Grebille~\cite{Mf-Struct-473K} solved the crystallographic structure
in the high temperature phase. They showed that the transition is
due to a rearrangement of the central $\rm CoO$ layer of the $\rm
Ca_2CoO_3$ subsystem. This rearrangement acts on the $\rm CoO_2$
layers through the O--Ca interaction between the two subsystems. It
results in a modification of both the cobalts and oxygens modulations
(but not of the average structure) along the $\vec a$ and $\vec c$
directions. While such a weak structural modification can account for
the metal-semiconductor transition (due to the large correlation
strength and associated closeness to the Mott transition), it cannot
account for the associated increase of the Seebeck
coefficient. Indeed, such an increase necessitate the opening of new
degrees of freedom at the Fermi level and the weakness of the
structural modifications within the $\rm CoO_2$ layers cannot account
the opening of either the higher cobalt spin states or the $e'_g$
orbitals. This study however show us that there is another degree of
freedom up to now ignored that should be taken into account, that is
the partial charge localization due to the action of the subsystem
modulation over the Fermi level orbital energies. Indeed a
modification of the modulation vector will be expected to modify the
energy profile pictured in figure~\ref{fig:delta}, thus the charge
localization configuration and the lattice pattern seen by the Fermi
electrons at the given temperature. Out of any other energetically
compatible hypothesis, the latter seems the most plausible.

Let us now extend our discussion to the other important parameters for the
thermoelectric performance. As we all know, a good figure of merit is
associated with a large Seebeck coefficient $S$, a low resistivity $\rho$ and
a low thermal conductivity $\kappa$.
\begin{itemize}
\item[$\bf S$~: ] We saw in the preceeding paragraphs that the Heikes formula
  is fully justified to describe the Seebeck coefficient of the layered cobalt
  oxides (provided a large $U/t$ value and a partial charge localization). The
  Heikes formula is a pure entropic one and only depends on the band filling
  of the $a_{1g}$ cobalts magnetic orbitals of the $\rm CoO_2$ layer.  Thus,
  as far as $S$ is concerned, the misfit character essentially acts (i) on the
  filling of the $\rm CoO_2$ layer, since the latter should adjust to the
  incommensurate ratio ($b_{\rm CoO_2}/b_{RS}$) in order to preserve the
  electro-neutrality of the system, (ii) on the renormalization of the
  $a_{1g}$ band width (through a partial localization). In both cases these
  are indirect effects (acting through the electrostatic interaction between
  the two layers) and the effect of the specific modulation vectors, or even
  their incommensurate character, will be expected to modulate the $S$ value,
  but not its order of magnitude.

\item[$\bf \rho$~:] In these $\rm CoO_2$-based systems, the observed low
  resistivity is essentially due to the triangular pattern of the transfer
  interactions in the $\rm CoO_2$ layer. Indeed, in a triangular lattice, the
  Mott transition is repelled at very high correlation values, and the system
  remains essentially metallic (as experimentally observed in the misfit
  cobalt oxides which present low metal to insulator transition temperatures,
  between 50K-80K for the present compound~\cite{Mf-TEP1,Mf-TEP2,Ca3Co4O9_3},
  around 20K for the $\rm [Sr_2Co_{1-x}Bi_xO_3][CoO_2]_{1.8}$
  compounds~\cite{BiSrCoO} or the
  $\rm[Bi_{1.95}Ba_{1.95}Rh_{0.1}O_4][RhO_2]_{1.8}$~\cite{BiBaRhO}, or even
  remaining metallic at all temperatures such as the $\rm
  [Sr_2O_2]_{0.53}[CoO_2]$~\cite{SrO}). As a matter of illustration, let us
  compare the $U/t\simeq 15$~\cite{Tr-Mott} ratio found for the metal to
  insulator transition of a half-filled, triangular, Hubbard model, and the
  $U/t= 0$ ratio for the same transition when the transfer interactions
  pattern is square. The misfit character can thus be expected to be non
  relevant compared to the basic triangular interaction pattern as far as the
  resistivity is concerned. At a higher order of precision, however, it is
  expected to sightly increases the resistivity, through the induced partial
  localization.
  
\item[$\bf \kappa$~:] The main effect of the misfit character is most probably
  on the thermal conductivity. Indeed, one can expect that the non periodicity
  and the associated modulated potential seen by otherwise equivalent ions,
  induce a partial localization of the phonons dispersion and thus a lowering
  of the thermal conductivity. This point is however difficult to quantify
  since it would necessitate a comparative study of the phonons dispersion
  spectra for different misfit and non misfit $\rm
  CoO_2$-based compounds. \\ 
\end{itemize} 

\section{Conclusion}

We performed ab initio calculations of the effective local parameters of the
thermoelectric $\rm Ca_3Co_4O_9$ compound, taking into account the
incommensurate distortions. As in the $\rm Na_x CoO_2$ family, cobalts
$a_{1g}$--$e'_g$ orbitals splitting is large so that the low energy physics
can be accounted for using the $a_{1g}$ orbitals only.  We showed that the
incommensurate modulation of the $\rm CoO_2$ layer cannot be neglected in
order to understand the compound physical properties. Indeed, site energies,
NN exchange, NN hopping as well as first neighbor repulsions are strongly
modulated due to the misfit character of the system. It particular the site
energies large modulation let us expect a partial Fermi electrons
localization. The effective exchange remains weak in amplitude, however
fluctuates between antiferromagnetic coupling and ferromagnetic one. This
results can be compared with the on-plane AFM-FM transition line in the
$\rm Na_xCoO_2$ family associated at $T=0\,\rm K$ with similar cobalt valency as
found in the present compound. 

We discussed the temperature behavior of the Seebeck coefficient and showed
that the cobalts $a_{1g}$ orbitals bandwidth is strongly renormalized, due to
very strong electron correlation ($U/t\sim 26$), and the modification of the
effective lattice pattern seen by the itinerant electrons, associated with the
partial charge localization. This strong bandwidth renormalization is most
probably responsible for the plateau due to the high temperature behavior
observed in the Seebeck coefficient on the $[150\,{\rm K},400\,{\rm K}]$
range. We further showed that he sudden increase of the TEP for
$T>T_c=396\,\rm K$, cannot be attributed either to a cobalt spin state change
or to a modification of the $a_{1g}$ vs $e'_g$ orbitals occupation. In fact
the 396\,K metal-semiconductor transition is associated with a structural
rearrangement within the $\rm Ca_2CoO_3$ subsystem inducing a modulation
modification within the $\rm CoO_2$ layer.  This modification induces a site
energy pattern change, and thus a localization pattern change, which we think
to be the best candidate for the change in the Seebeck behavior.

\acknowledgments The authors thank Prof. Dominique Grebille and Dr. Olivier
Perez for helpful discussion on the crystallographic aspects, as well as
Dr. Daniel Maynau for providing us with the CASDI suite of programs. These
calculations where done using the CNRS IDRIS computational facilities under
project n$^\circ$1842 and the CRIHAN computational facilities under project
n$^\circ$2007013.


\end{document}